\documentclass[11pt]{article}
\usepackage{amsmath}
\usepackage{amssymb}
\usepackage{amsfonts}
\usepackage{amscd}
\usepackage{accents}
\usepackage{pst-all}
\usepackage{epsfig}
\usepackage{graphicx,amssymb,amsmath,wasysym}
\date{\today}
\newcommand{\bmat}{\left(\begin{array}}
\newcommand{\emat}{\end{array}\right)}
\newcommand{\be}{\begin{equation}}
\newcommand{\ee}{\end{equation}}
\newcommand{\ba}{\begin{eqnarray}}
\newcommand{\ea}{\end{eqnarray}}

\def\lsim{\raise0.3ex\hbox{$\;<$\kern-0.75em\raise-1.1ex\hbox{$\sim\;$}}}
\def\gsim{\raise0.3ex\hbox{$\;>$\kern-0.75em\raise-1.1ex\hbox{$\sim\;$}}}

\def\be{\beta}

\usepackage{graphicx}
\usepackage[tight]{subfigure}

\begin{document}
\vspace*{-.6in} \thispagestyle{empty}
\begin{flushright}
DESY 11-194
\end{flushright}
\baselineskip = 20pt

\vspace{.5in} {\Large
\begin{center}
{\bf Higgs--induced lepton flavor  violation}

\end{center}}

\vspace{.5in}

\begin{center}
{\bf  Andreas Goudelis$^1$, Oleg Lebedev$^1$  and  
Jae-hyeon Park$^2$ }

\vspace{.5in}

\emph{$^1$DESY Theory Group, 
Notkestra{\ss}e 85, D-22607 Hamburg, Germany \\
$^2$Institut f\"{u}r Kern- und Teilchenphysik,
TU Dresden, 01069 Dresden, Germany}

\end{center}

\vspace{.5in}

\begin{abstract}
Due to the smallness of the lepton Yukawa couplings,
higher--dimensional operators can give a significant contribution
to the lepton masses. In this case,
the lepton mass matrix and the matrix of lepton--Higgs 
couplings are misaligned  leading to
lepton flavor violation (LFV)    mediated by the Standard Model Higgs boson. 
We derive model--independent bounds on the Higgs flavor violating 
couplings and 
 quantify LFV   in   decays of leptons 
and electric dipole moments 
for a class of  lepton--Higgs  operators contributing to lepton masses. 
We find significant Higgs--mediated LFV effects at both 1--loop and 2--loop 
levels. 
\end{abstract}

\noindent

\newpage

\section{Introduction}

The flavor puzzle of the Standard Model remains one of the outstanding problems
in particle physics. Masses of elementary fermions range over many orders of magnitude, 
forming a pattern which cannot be explained within the Standard Model.
Also, it is not known whether these masses are generated by renormalizable terms in the
Lagrangian or higher order operators also contribute  \cite{Babu:1999me,Giudice:2008uua}. 
This issue will be studied at the LHC by means of  the Higgs couplings measurements \cite{Lafaye:2009vr},
which will give us  at least partial answers. 
Complementary information about possible effects of non--renormalizable operators
is provided by  observables sensitive to flavor violation.

In this work, we focus on the lepton sector due to its particular sensitivity to flavor violation.
We will consider a class of higher dimensional operators involving the Higgs field which 
affect the lepton masses.
Although such operators are suppressed by a ``new physics'' scale $M$, they can still give a significant
contribution to the lepton masses due to the smallness of the lepton Yukawa couplings. 
In particular, we will focus on operators of the type 
\begin{equation} 
-\Delta {\cal L}=  Y_{ij}^{(0)} ~H~ \bar L_{Li} l_{Rj} + 
  Y_{ij}^{(1)} {H^\dagger H \over M^2}    ~H~ \bar L_{Li} l_{Rj} + ~...
~+~ {\rm h.c.} \;,
\end{equation}
as well as higher order operators.
Here $Y_{ij}^{(0)} $ and $Y_{ij}^{(1)} $ are {\it a priori} independent flavor matrices.
Since the resulting lepton mass matrix and the matrix of the Higgs couplings are in general
misaligned in flavor space, the presence of higher order operators leads to 
flavor violation mediated by the Standard Model Higgs boson. 
The rotation to the mass eigenstate basis may involve large angles, as hinted by the neutrino sector,
which amplifies the effect. For example, the processes involving light generations such as $\mu \rightarrow e \gamma$ can be 
dominated by loops involving the
$\tau$ lepton. 
The existing bounds on LFV observables
then place strict constraints on the scale of new physics and/or the type of admissible Yukawa
textures. 

Some aspects of lepton flavor violation in Higgs interactions have been considered before
(for early studies, see \cite{Bjorken:1977vt,McWilliams:1980kj,Shanker:1981mj}),
although a systematic study is lacking.
For example, LFV Higgs decays  $h \rightarrow l_i l_j$  
were analyzed in  \cite{DiazCruz:1999xe,Han:2000jz}. 
The Higgs--mediated decay $\tau \rightarrow \mu\gamma$ was considered in \cite{Aranda:2008si},
where it was found that this mode is not particularly constraining. Effects of higher
order $H^\dagger H$--operators on neutrino masses and LFV in a 2 Higgs doublet model were
studied in \cite{Bonnet:2009ej}. 
Finally, Ref.~\cite{Buras:2010mh} summarizes   Higgs--induced quark FCNC effects in  
extended Higgs models.

The paper is organized as follows. In Section 2, we derive model--independent  bounds on the 
 flavor violating  couplings of the Higgs boson. In Section 3.1, we apply these bounds to the
case of dimension 6 operators and obtain constraints on the scale of new physics and 
on the rotation angles. In Section 3.2, we study the possibility that the lepton mass hierarchy
is created entirely by non--renormalizable operators. In Section 4, we present our conclusions.

\section{Bounds on the Higgs couplings}

\begin{figure}
  \centering
  \subfigure[]{\includegraphics[height=7em]{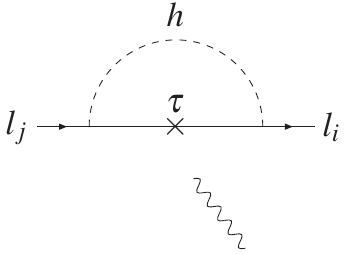}}
  \hspace{0.5in}
  \subfigure[]{\includegraphics[height=7em]{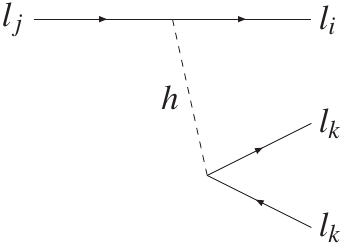}}
  \caption{Leading  Higgs--mediated   contributions to the lepton dipole operators and
     3--body decays.  }
  \label{fig:diagrams}
\end{figure}

The relevant Lagrangian describing interactions of the physical Higgs boson $h$ with leptons
is given by
\begin{equation}
\Delta {\cal L} = -{y_{ij}\over \sqrt{2}} ~h ~\bar l_i P_R l_j ~+~ {\rm h.c.} \;,
\label{Lhiggs}
\end{equation}
where $P_R = (1+\gamma_5)/2$. This interaction induces the following 
flavor changing dipole operator (Fig.~\ref{fig:diagrams}), 
\begin{equation}
\label{eq:dipole ops}
{\cal L}_{\rm eff_1}=e L_{ij} ~\bar l_i \sigma^{\mu\nu} F_{\mu\nu}P_L l_j
~+~ {\rm h.c.} \;,
\end{equation}
where $\sigma^{\mu\nu}={i\over 2}  [\gamma^\mu, \gamma^\nu]$,
$P_L = (1-\gamma_5)/2$,
and 
\begin{equation}
L_{ij}=  {y_{3i}^* y_{j3}^* \over 64 \pi^2 m_h^2}~ m_\tau \ln {m_\tau^2
\over m_h^2} \;. 
\end{equation}
Here we have kept only the leading  in $m_\tau /m_h$
$\tau$--lepton contribution in the loop since the others are negligible
for our purposes.
The corresponding branching ratio is given by
\begin{equation}
{\rm BR}(l_j \rightarrow l_i \gamma) = 
{\rm BR}(l_j \rightarrow l_i \nu_j \bar \nu_i) \times
{192 \pi^3 \alpha \over G_F^2 m_j^2 } ~\bigl( 
\vert L_{ij}\vert^2 + \vert L_{ji}\vert^2
\bigr) \;.
\end{equation}
The Higgs interactions also induce flavor--diagonal 
dipole operators at one loop:
\begin{equation}
{\cal L}_{\rm eff_2}
=e~ {\rm Re} L_{ii} ~\bar l_i \sigma^{\mu\nu} F_{\mu\nu} l_i ~-~
i e~ {\rm Im} L_{ii} ~\bar l_i \sigma^{\mu\nu} F_{\mu\nu}\gamma_5 l_i 
~+~ {\rm h.c.} \;.
\end{equation}
These are constrained by the charged lepton anomalous magnetic moments  and 
electric dipole moments,
\begin{eqnarray} 
\vert \delta a_i \vert &=& 4 m_i \big\vert {\rm Re} L_{ii} \big\vert
\;, \nonumber\\
\vert  d_i \vert &=& 2 e \big\vert {\rm Im} L_{ii} \big\vert \;.
\end{eqnarray}

Finally, there are tree--level processes 
$l_j \rightarrow l_i l_k l^+_k$
induced by (\ref{Lhiggs}). Their branching ratio is given by
\begin{equation}
{\rm BR}(l_j \rightarrow l_i l_k l^+_k)= 
{\rm BR}(l_j \rightarrow l_i \nu_j \bar \nu_i) \times
{(4 - \delta_{ik}) \over 256 G_F^2 m_h^4}~
\vert y_{kk}   \vert^2 
\big( \vert y_{ij}\vert^2 + \vert y_{ji}\vert^2     \big)\;.
\end{equation}
The resulting bounds on the Higgs couplings are presented in Table 1.
Most  experimental constraints are taken   from Particle Data Group \cite{PDG},
while the recent bound on BR$(\mu \rightarrow e \gamma)$ is from   \cite{MEG2011}
and the interpretation of $\delta a_e$ is due to \cite{Girrbach:2009uy}.
 The third column shows
the combination of Higgs couplings constrained by a particular observable,
while the fourth one shows representative bounds on the couplings under
the assumptions $y_{ij}=y_{ji}$ and $y_{ii}=m_i/v$, 
as in the Standard Model.\footnote{Here we have not included the bound from 
$\mu$ to $e$ conversion on nuclei since it depends on the Higgs couplings
to light quarks, which are uncertain in our framework. In any case,
for the SM Higgs--quark couplings, the bound is of order
$\sqrt{   \vert y_{12} \vert^2 + \vert y_{21} \vert^2    } < 5\times 10^{-4}$
for $m_h=200$ GeV,
which is superceded by the 2--loop  Barr--Zee bound to be studied below.}

\begin{table}[htbp]
\begin{center}
\begin{tabular}{|l|c|c|l|}
\hline 
 observable &  present  limit &
  constraint &  constraint for     \cr
 & & & $y_{ij}=y_{ji}~,~ y_{ii}=m_i/v$     \cr
& & & \\
\hline
$\mbox{BR}\, (\mu \rightarrow e \gamma)$ &
  $2.4 \times 10^{-12}$   & $\Big(\vert y_{31} y_{23}\vert^2 +
\vert y_{32} y_{13}\vert^2 \Big)^{1/4}< 7\times 10^{-4}$  &  $\sqrt{\vert
y_{13}
y_{23} \vert} <6\times 10^{-4}$  \\
$\mbox{BR}\, (\tau \rightarrow \mu \gamma)$  
  & $4.4 \times 10^{-8}$  & $ \Big( \vert y_{33}\vert^2~ ( \vert y_{32}\vert^2
+ \vert y_{23}\vert^2  )
\Big)^{1/4} < 5\times 10^{-2}$  & $ \vert y_{23}\vert < 2\times 10^{-1}$  \\
$\mbox{BR}\, (\tau \rightarrow e \gamma)$ &  
  $3.3 \times 10^{-8}$   &  $ \Big( \vert y_{33}\vert^2~ ( \vert y_{31}\vert^2
+ \vert y_{13}\vert^2  )
\Big)^{1/4} < 5\times 10^{-2}$   & $ \vert y_{13}\vert < 2\times 10^{-1}$   \\
$\mbox{BR}\, (\mu \rightarrow e e e)$ &  
  $1.0 \times 10^{-12}$   &  $\Big( \vert y_{11}\vert^2~ ( \vert y_{21}\vert^2
+ \vert y_{12}\vert^2  )
\Big)^{1/4} < 2\times 10^{-3}$    &  
$ \vert y_{12}\vert < 1$     \\
$\mbox{BR}\, (\tau \rightarrow \mu \mu \mu)$ &   
  $2.1 \times 10^{-8}$ &   $ \Big( \vert y_{22}\vert^2~ ( \vert y_{23}\vert^2
+ \vert y_{32}\vert^2  )
\Big)^{1/4} < 4\times 10^{-2}$    &  $ \vert y_{23}\vert < 1.7$   \\
$\mbox{BR}\, (\tau \rightarrow e e e)$ &   
  $2.7 \times 10^{-8}$ &   $ \Big( \vert y_{11}\vert^2~ ( \vert y_{13}\vert^2
+ \vert y_{31}\vert^2  )
\Big)^{1/4} < 4\times 10^{-2}$    &  $ \vert y_{13}\vert < {\cal O}(10^2)$   \\
$\mbox{BR}\, (\tau \rightarrow e \mu \mu)$ &   
  $2.7 \times 10^{-8}$ &   $ \Big( \vert y_{22}\vert^2~ ( \vert y_{13}\vert^2
+ \vert y_{31}\vert^2  )
\Big)^{1/4} < 4\times 10^{-2}$    &  $ \vert y_{13}\vert < 1.7$   \\
\hline
$d_e$ (e$\cdot$cm) & $1.1 \times 10^{-27}$  &  
$\sqrt{ \big\vert{\rm Im}(y_{31} y_{13})\big\vert } < 2\times 10^{-4} $  &
$\sqrt{ \big\vert{\rm Im}( y_{13}^2)\big\vert } < 2\times 10^{-4} $
     \\
$d_\mu$ (e$\cdot$cm) & $ 3.7 \times 10^{-19}$ & 
$\sqrt{ \big\vert{\rm Im}(y_{32} y_{23})\big\vert } < 4.1 $ 
     &  $\sqrt{ \big\vert{\rm Im}( y_{23}^2)\big\vert } < 4.1 $    \\
 $\delta a_e$  & $2.3\times 10^{-11}$  &
 $\sqrt{ \big\vert{\rm Re}(y_{31} y_{13})\big\vert } < 0.14 $   & 
$\sqrt{ \big\vert{\rm Re}( y_{13}^2)\big\vert } < 0.14 $ \\
 $\delta a_\mu$  & $40\times 10^{-10}$   &
 $\sqrt{ \big\vert{\rm Re}(y_{32} y_{23})\big\vert } < 0.13 $   & 
$\sqrt{ \big\vert{\rm Re}( y_{23}^2)\big\vert } < 0.13 $ \\
\hline
\end{tabular}
\caption{Current experimental limits on  flavor and CP violating
observables in the lepton sector, and the corresponding constraints on the Higgs couplings. The displayed bounds on $y_{ij}$ correspond to $m_h=200$ GeV;
for other Higgs masses they are to be multiplied by $m_h/(200 $ GeV).   }
\label{table1}
\end{center}
\end{table}

\subsection{2--loop contributions}

\begin{figure}
  \centering
  {\includegraphics[height=9em]{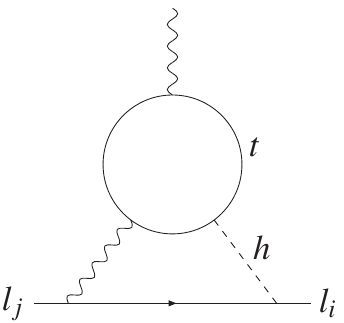}}
  \caption{Leading  2--loop Higgs--mediated  contribution to the lepton dipole operators.  }
  \label{fig:diagram-2}
\end{figure}

\begin{table}[htbp]
\begin{center}
\begin{tabular}{|c|c|}
\hline 
 observable &  
  constraint   \\
\hline
$\mbox{BR}\, (\mu \rightarrow e \gamma)$ & $\Bigl(  \vert y_{12} \vert^2 + \vert y_{21} \vert^2    \Bigr)^{1/2} < 6\times 10^{-6}$ \\
$\mbox{BR}\, (\tau \rightarrow \mu \gamma)$ & $\Bigl(  \vert y_{23} \vert^2 + \vert y_{32} \vert^2    \Bigr)^{1/2} < 4\times 10^{-2}$ \\
$\mbox{BR}\, (\tau \rightarrow e \gamma)$ & $\Bigl(  \vert y_{13} \vert^2 + \vert y_{31} \vert^2    \Bigr)^{1/2} < 3\times 10^{-2}$ \\
$d_e$ &  $\bigl\vert {\rm Im }~ y_{11}  \bigr\vert < 6 \times 10^{-7}$    \\
$d_\mu$ &  $\bigl\vert {\rm Im }~ y_{22}  \bigr\vert < {\cal O}(10^2)  $    \\
$\delta a_e$ &  $\bigl\vert {\rm Re }~ y_{11}  \bigr\vert < 0.27$ \\
$\delta a_\mu$ &  $\bigl\vert {\rm Re }~ y_{22}  \bigr\vert < 0.21$ \\
\hline
\end{tabular}
\caption{ Two loop constraints for $m_h=200$ GeV. 
(The bounds get tighter  by a factor of 1.5 for $m_h=100$ GeV.)  }
\label{table2}
\end{center}
\end{table}

Two loop diagrams of certain types can be important as they are
suppressed by only one power of small Yukawa couplings \cite{Bjorken:1977vt}.
In what follows, we include the leading  Barr--Zee type diagrams \cite{Barr:1990vd}
with  the top quark in the loop  (Fig.~\ref{fig:diagram-2})     \cite{Chang:1993kw}.  
Since the coupling of the top quark to the Higgs is essentially unaffected
by the higher dimensional operators, we will assume the SM value $y_{tt}=1$.
Then \cite{Chang:1993kw},
\begin{equation}
{\rm BR}(l_j \rightarrow l_i \gamma) = 
{\rm BR}(l_j \rightarrow l_i \nu_j \bar \nu_i) \times 
{8 \alpha^3 v^2 \over 3 \pi^3 m_j^2} ~ f^2(z)~ \bigl(   
\vert y_{ij} \vert^2 + \vert y_{ji} \vert^2   \bigr) \;,
\end{equation}
where $v=$ 174 GeV and 
\begin{equation}
f(z)= {1\over 2} z \int_0^1 dx { 1-2x(1-x) \over x(1-x) -z  } \ln 
{x(1-x) \over z} \;,
\end{equation}
with $z= m_t^2 /m_h^2$. 
The resulting constraints are presented in Table 2. Since $f(z)$ varies from 0.76 
for $m_h=200$ GeV to 1.14 for $m_h=100$ GeV, these bounds are not very sensitive to 
the Higgs mass. 

The flavor diagonal transitions can also be extracted from the amplitude
computed in \cite{Chang:1993kw}. We find
\begin{equation}
L_{ii}= {  \alpha \over 24 v  \pi^3   }~ f(z)~y_{ii}^* \;.
\end{equation}
The EDMs and  $g-2$ then constrain the diagonal couplings as summarized in Table 2.
We note that the bounds from BR$(\mu \rightarrow e \gamma)$ and $d_e$ are particularly
important.\footnote{In 2 Higgs doublet models, Barr--Zee contributions 
often dominate  at low $\tan\beta$ \cite{arXiv:1001.0434}.}

\section{Lepton flavor violation from Yukawa--type interactions}

\subsection{Dimension--6 operators}

The lepton sector is particularly sensitive to BSM flavor structures
due to the smallness of the lepton masses and strong constraints on 
lepton flavor violation. Since the lepton
Yukawa couplings in the SM can be as small as $10^{-5}$, 
higher dimensional operators  involving the Higgs field 
 can give a significant contribution to the lepton masses.
In this section we consider an effect of dimension 6 operators of this sort,
\begin{equation} 
-\Delta {\cal L}=  H~ \bar L_{Li} l_{Rj} 
\left(  Y_{ij}^{(0)} +  Y_{ij}^{(1)} {H^\dagger H \over M^2}    \right) 
~+~ {\rm h.c.} \;,
\end{equation}
which amounts to replacing the constant SM Yukawa couplings with Higgs--dependent
ones,
\begin{equation}
 Y_{ij} =  Y_{ij}^{(0)} +  Y_{ij}^{(1)} {H^\dagger H \over M^2} \;.
\label{dim6-Y}
\end{equation}
Here $ Y_{ij}^{(0)} ,  Y_{ij}^{(1)} $ are in general independent flavor matrices
and $M$ is the new physics scale. An immediate consequence of the above 
Lagrangian is that the SM Higgs boson mediates tree level flavor changing
neutral currents. Indeed, the lepton mass matrix is given by
\begin{equation} 
M_{ij} = v \left(  Y_{ij}^{(0)} +  Y_{ij}^{(1)} { v^2 \over M^2}    \right) \;,
\end{equation}
whereas the matrix of couplings of the physical Higgs boson    is
\begin{equation}
{\cal Y}_{ij}= Y_{ij}^{(0)} +  3 Y_{ij}^{(1)} { v^2 \over M^2} \;, 
\end{equation}
where we have used the convention $H^0 = v + h/\sqrt{2}$. Clearly, these matrices are
in general misaligned in flavor space and the Higgs couplings in the mass eigenstate basis 
(cf. Eq.~\ref{Lhiggs}) can change flavor. The latter are  given by 
\begin{equation}
y = U_L^\dagger ~ {\cal Y} ~ U_R \;,
\end{equation}
where the unitary matrices $U_L , U_R$ diagonalize the lepton mass matrix,
$U_L^\dagger ~ M ~ U_R $=$ {\rm diag} (m_e, m_\mu, m_\tau)$.

To understand the constraints on the dim--6 operators, we  vary the 
proportion between $Y_{ij}^{(0)}$ and $Y_{ij}^{(1)} { v^2 / M^2} $
and scan over different $U_L , U_R$. We allow  for large angle rotations,
as hinted by the neutrino sector, while being agnostic about the neutrino
mass matrix which presumably is provided by the seesaw mechanism.
The resulting $y_{ij}$ are then subject to the
bounds described in the previous section.

For the scan, we use the following procedure.
The Yukawa textures are generated through 
\begin{equation}
  \label{eq:biunitary}
  Y = U_L ~  {1\over v} ~ \mathrm{diag} (m_e, m_\mu, m_\tau)     ~     U_R^\dagger  \;,
\end{equation}
by scanning over $U_L, U_R$. 
Modulo phase redefinitions
of the lepton fields,
$U_{L,R}$ can  be chosen as
\begin{equation}
  \label{eq:unitary matrix}
  U_L =  V_L \;, \quad
  U_R =  V_R ~\Theta \;,
\end{equation}
with unitary $V_{L,R}$ parametrized by  three mixing angles and a single phase
as in Refs.~\cite{PDG,Chau:1984fp}, and 
 $\Theta$ being  a diagonal phase matrix,
\begin{equation}
  \Theta = \mathrm{diag}(e^{i\phi_1}, e^{i\phi_2}, e^{i\phi_3}) \;,
\label{theta}
\end{equation}
subject to the constraint
$ \phi_1 + \phi_2 + \phi_3 = 0 . $ Note that there are 4 reparametrization--invariant
phases in the lepton mass matrix since 5 out of 9 original phases can be eliminated by the phase transformations of the left--handed  and right--handed leptons.  

Having generated $Y_{ij}$ (setting $H \rightarrow v$), we split it into
two pieces according to Eq.~\ref{dim6-Y}. Clearly, if $Y_{ij}^{(0)}$ dominates,
$M_{ij}$ and ${\cal Y}_{ij}$ are almost aligned, and  the lepton
FCNC are suppressed. The latter therefore set a bound on the allowed proportion 
of  $Y_{ij}^{(1)} v^2/M^2 $ in the lepton mass matrix. 
We scan over the following parameters:
\begin{itemize}
 \item $6$ angles of  $V_{L}$, $V_{R}$ 
 \item $2$ phases of $V_{L}$, $V_{R}$ and 2 phases of $\Theta$
 \item $9$ complex parameters  $Y_{ij}^{(0)}$
\end{itemize}
Note that, given  $Y_{ij}$ and $Y_{ij}^{(0)}$, 
the remaining piece  $Y_{ij}^{(1)} v^2/M^2$  is determined by Eq.~\ref{dim6-Y}.
In what follows we present our results for two cases: 
$Y_{ij}^{(0)}$ is varied first in the range $ [0.5,1]  \times  Y_{ij}$ and 
then  in the range $ [0.9,1]  \times  Y_{ij}$.\footnote{This range applies 
separately to the real and imaginary parts of $Y_{ij}^{(0)}$.}
 This restricts the relative
contribution of the dim--6 term to no more than 50\% and 10\%, respectively.

\begin{figure}
\begin{center}
\hspace{-1.3cm}
\includegraphics[width=8.5cm]{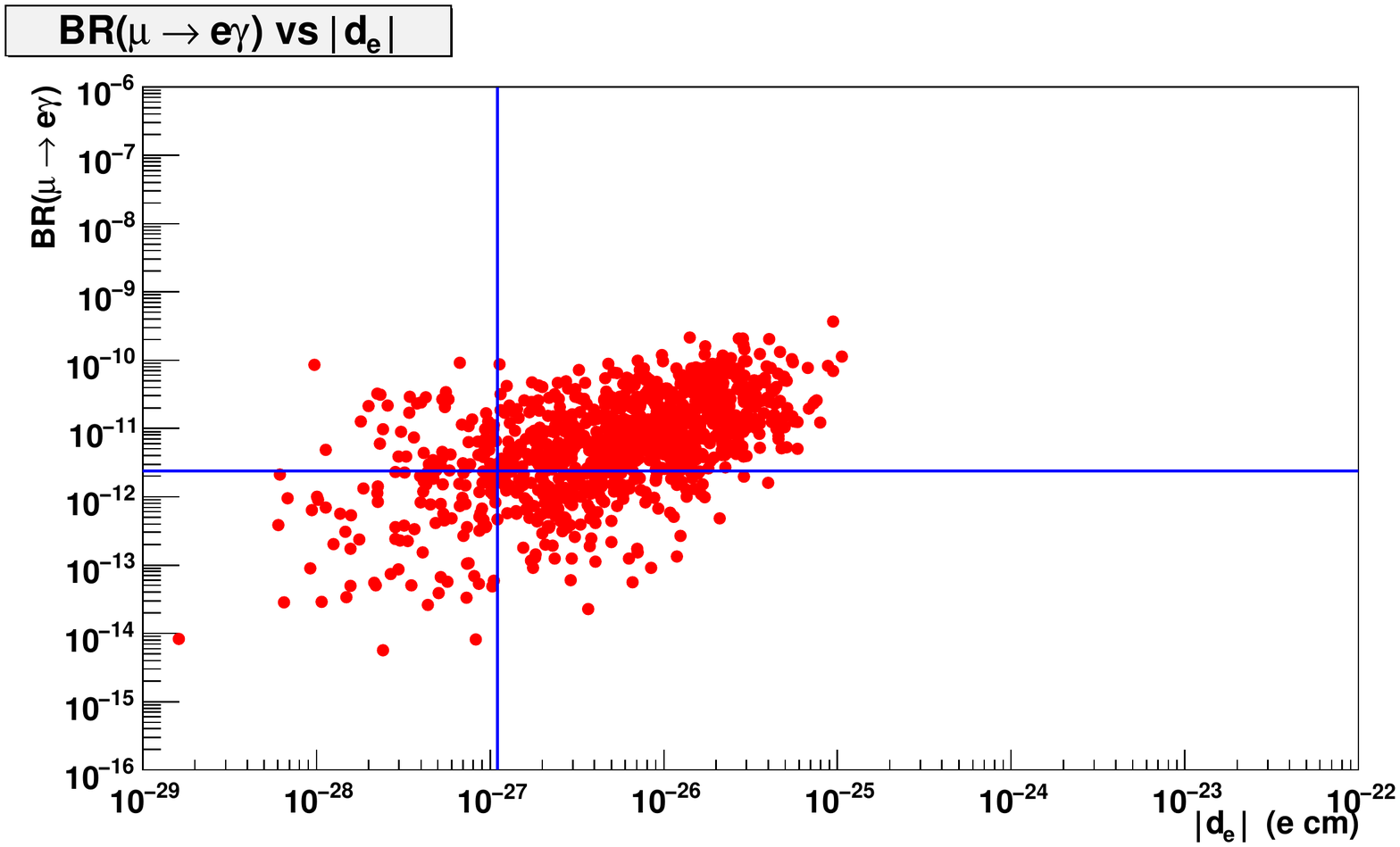}\hspace{0.01cm}
\includegraphics[width=8.5cm]{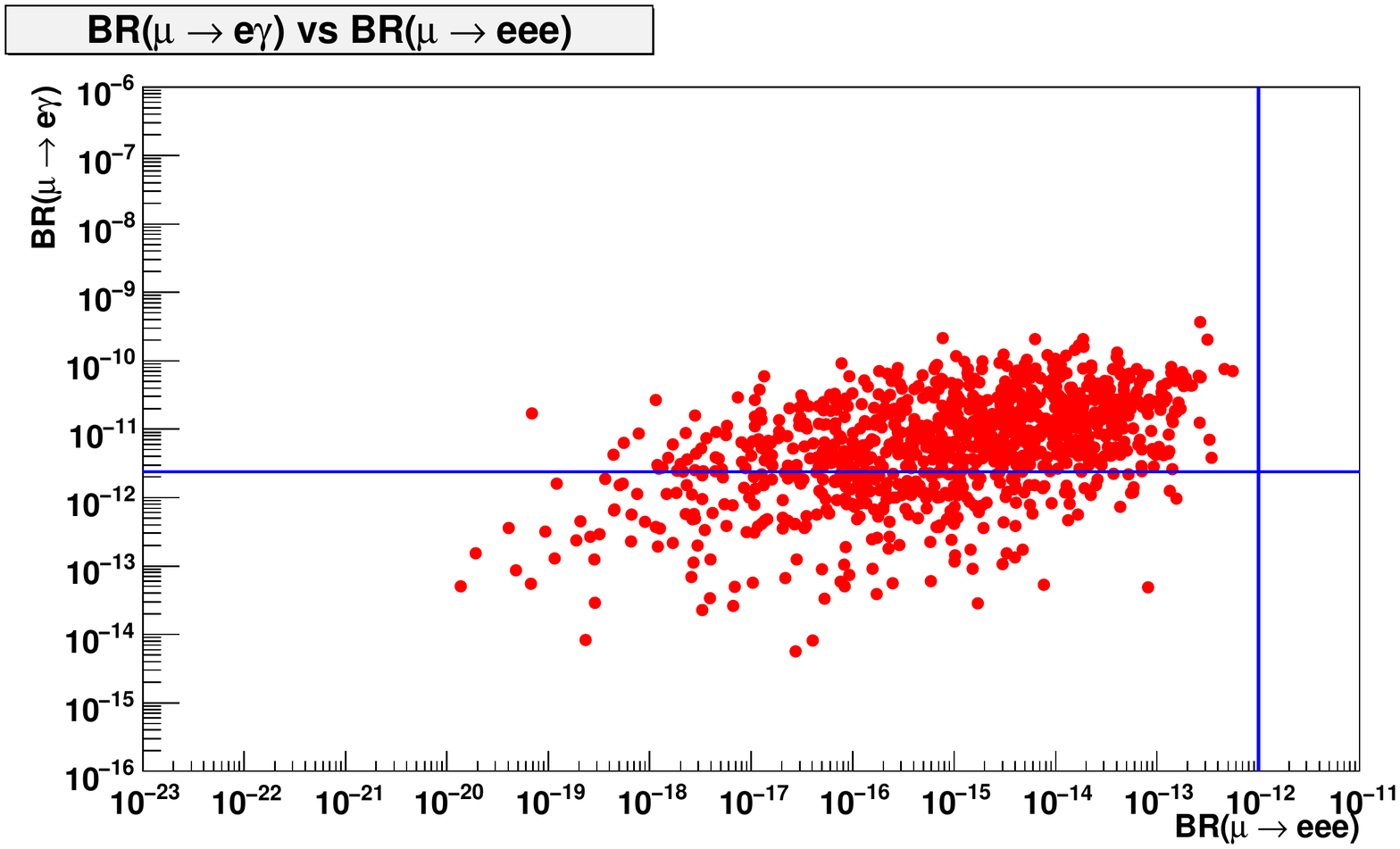}\\
\hspace{-1.3cm} 
\includegraphics[width=8.5cm]{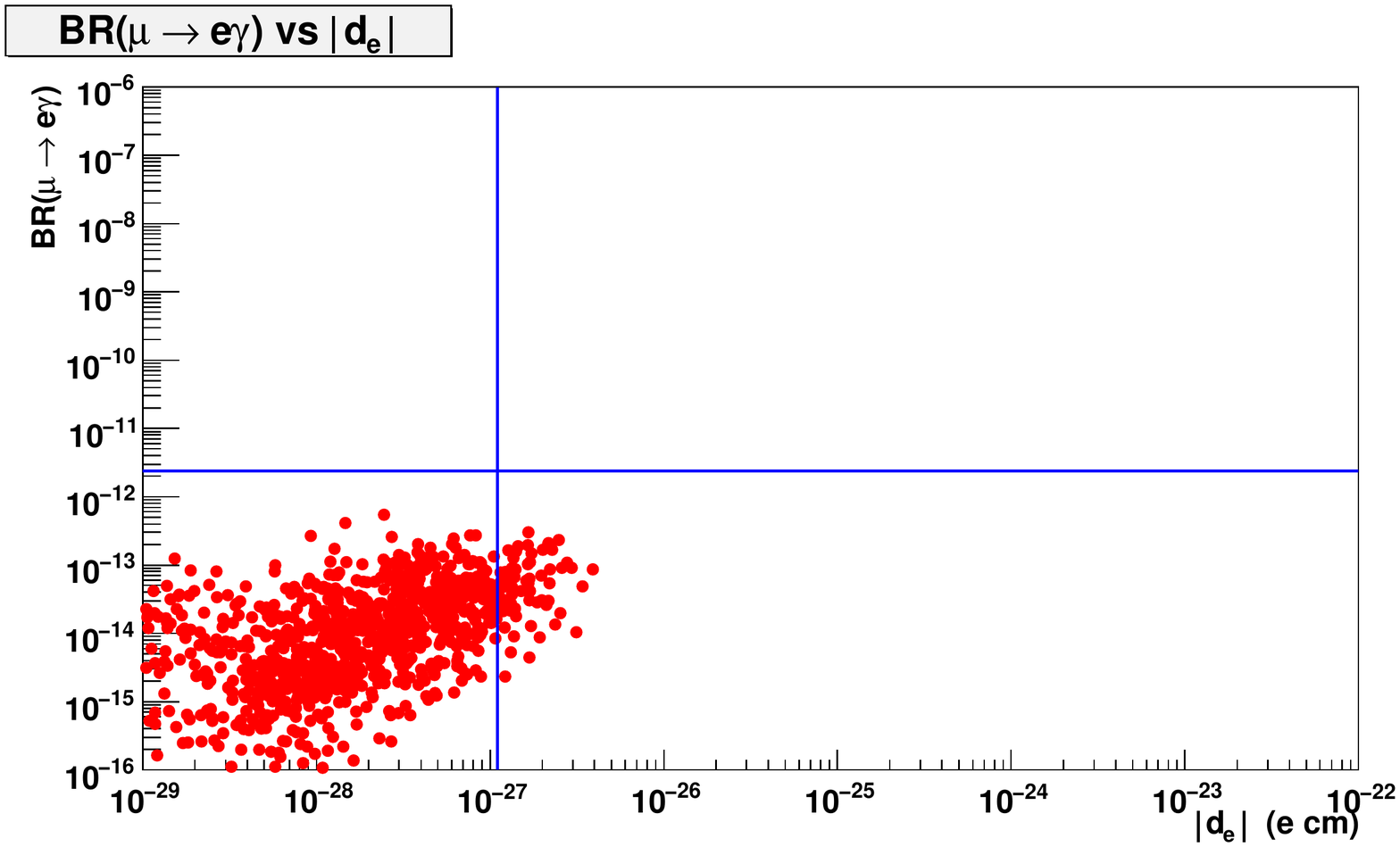}\hspace{0.01cm}
\includegraphics[width=8.5cm]{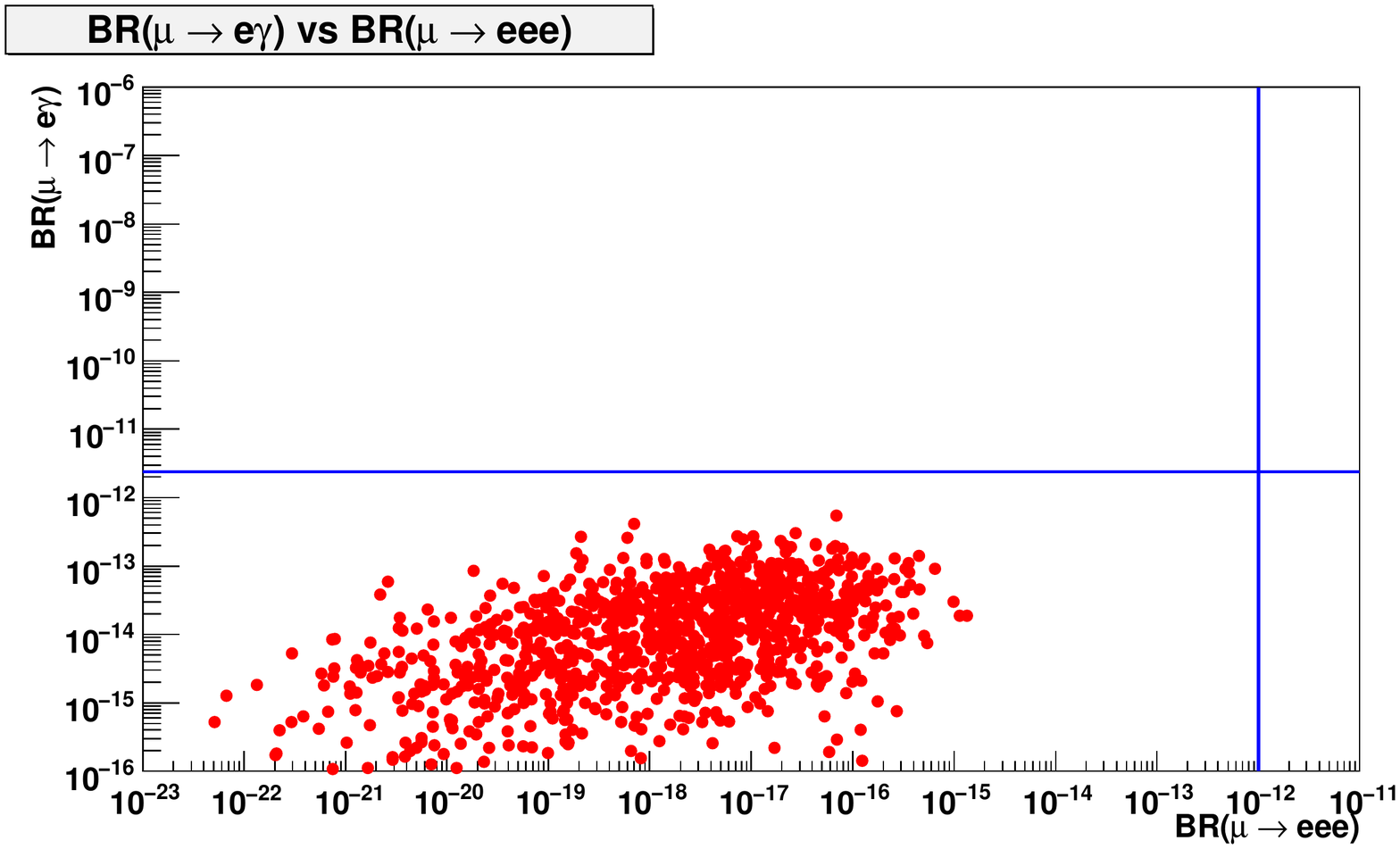}
\end{center}
\caption{$\mbox{BR} (\mu \rightarrow e \gamma)$ vs $\left| d_e \right|$
(left)
and $\mbox{BR} (\mu \rightarrow e \gamma)$ vs $\mbox{BR} (\mu
\rightarrow e e e)$ (right) for arbitrary rotation angles and $m_h=200
$ GeV.  Dimension--6 operators contribute up to 50\% (top) and 10\% (bottom)
to $Y_{ij}$. The experimental limits are given by the straight lines. }
\label{resdim6}
\end{figure}

Our results are presented in Fig.~\ref{resdim6}. We see that if the dim--6
operators are allowed to contribute as much as 50\% to $Y_{ij}$, 
BR$(\mu \rightarrow e \gamma)$ and $d_e$ are overproduced by up to 
2 orders of magnitude. On the other hand, if this contribution is below 
10\%, most points are allowed for $m_h = 200$ GeV.
Thus, allowing for arbitrary rotation angles, we find an empirical constraint
\begin{equation}
\biggl\vert  Y_{ij}^{(1)} { v^2 \over M^2} \biggr\vert
< 0.1  \vert Y_{ij} \vert  \times  {200~{\rm GeV}\over m_h } \;.
\label{const}
\end{equation}
This can be reinterpreted in terms of the bounds on the new physics scale 
$M$. For the two  limiting cases of similar $ Y_{ij}^{(1)}$ and $Y_{ij}$,
and   order one $ Y_{ij}^{(1)}$, we get 
\begin{eqnarray}
   Y_{ij}^{(1)} \sim  Y_{ij} & \Rightarrow & M > 500~{\rm GeV} \times {200~{\rm GeV}\over m_h } ~, \nonumber\\
  Y_{ij}^{(1)} \sim 1 & \Rightarrow & M > 200~{\rm TeV} \times {200~{\rm GeV}\over m_h } ~.
\label{interp}
\end{eqnarray}
In the latter case, we used the most restrictive Yukawa couplings 
${\cal O}(10^{-5}$)  
involving the electron. Let us note that we find  observables
other than BR$(\mu \rightarrow e \gamma)$ and $d_e$  
far less constraining  and also confirm numerically that the diagrams
with muons/electrons in the loop are unimportant.
The latter statement is easy to understand.
For the $\tau$ contribution,  the relevant Higgs vertices 
involve the $\tau$ mass times a mixing parameter, e.g.
\begin{equation} 
y_{13} \sim  \epsilon_{13} ~ {m_\tau \over v} ~,
\end{equation}
and, in addition, the mass insertion in the loop is $m_\tau$. This enhances
the $\tau$ loop  by orders of magnitude. The couplings 
involving only the first two generations are typically bounded by $m_\mu/v 
\sim  10^{-3}$, unless there is a very strong mixing with the $\tau$. 
To account for the muon loop contribution to  BR$(\mu \rightarrow e \gamma)$ and $d_e$, 
 one can simply  rescale 
the bounds of Table 1 by $\sqrt{m_\tau/ m_\mu}$ and replace index 3 by 2.
Keeping in mind that there are   mixing angles appearing at the vertices
so that the actual couplings are smaller than $10^{-3}$, one finds that 
these constraints are   satisfied automatically.

\begin{figure}
\begin{center}
\hspace{-1.3cm}
\includegraphics[width=8.5cm]{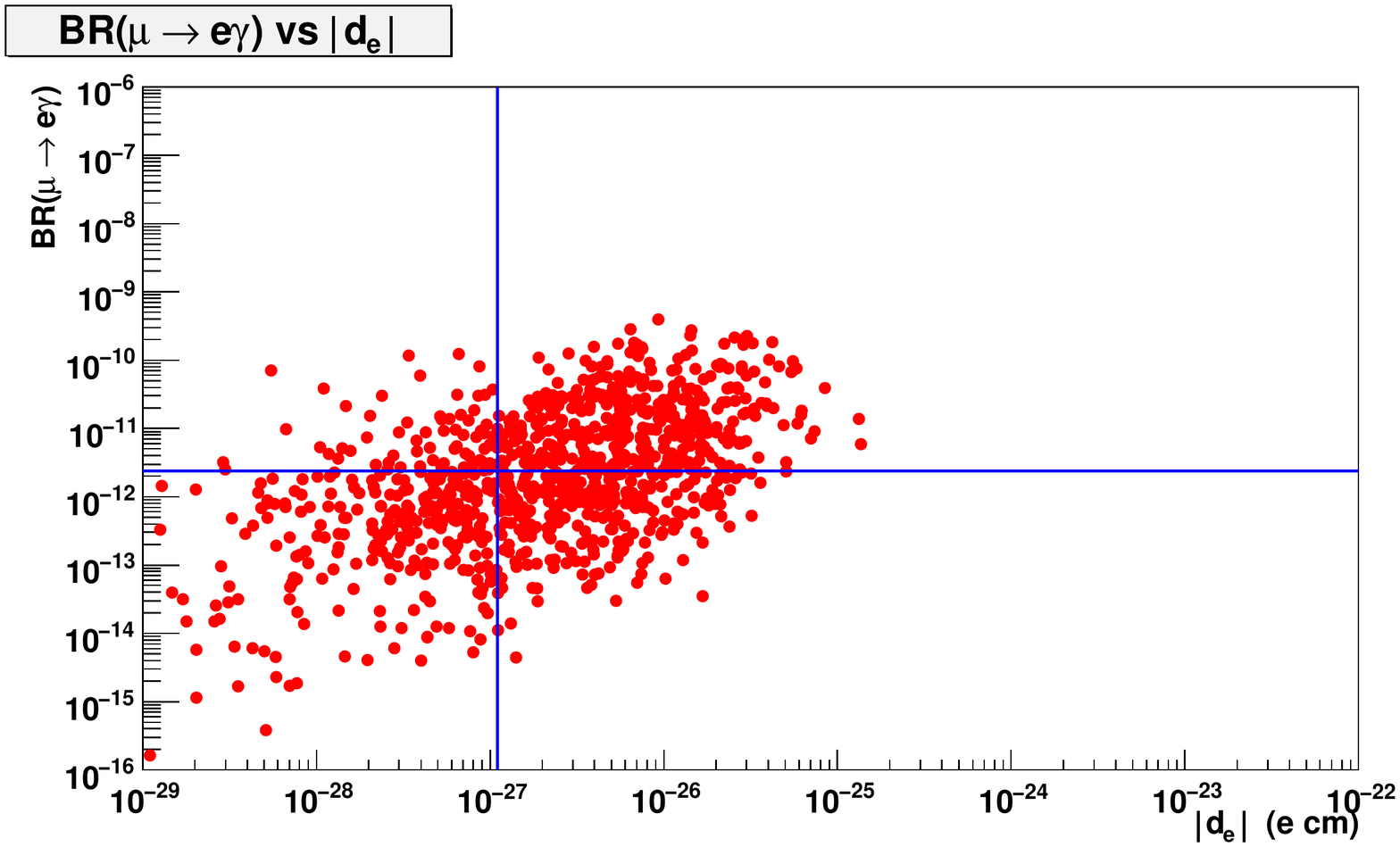}\hspace{0.01cm}
\includegraphics[width=8.5cm]{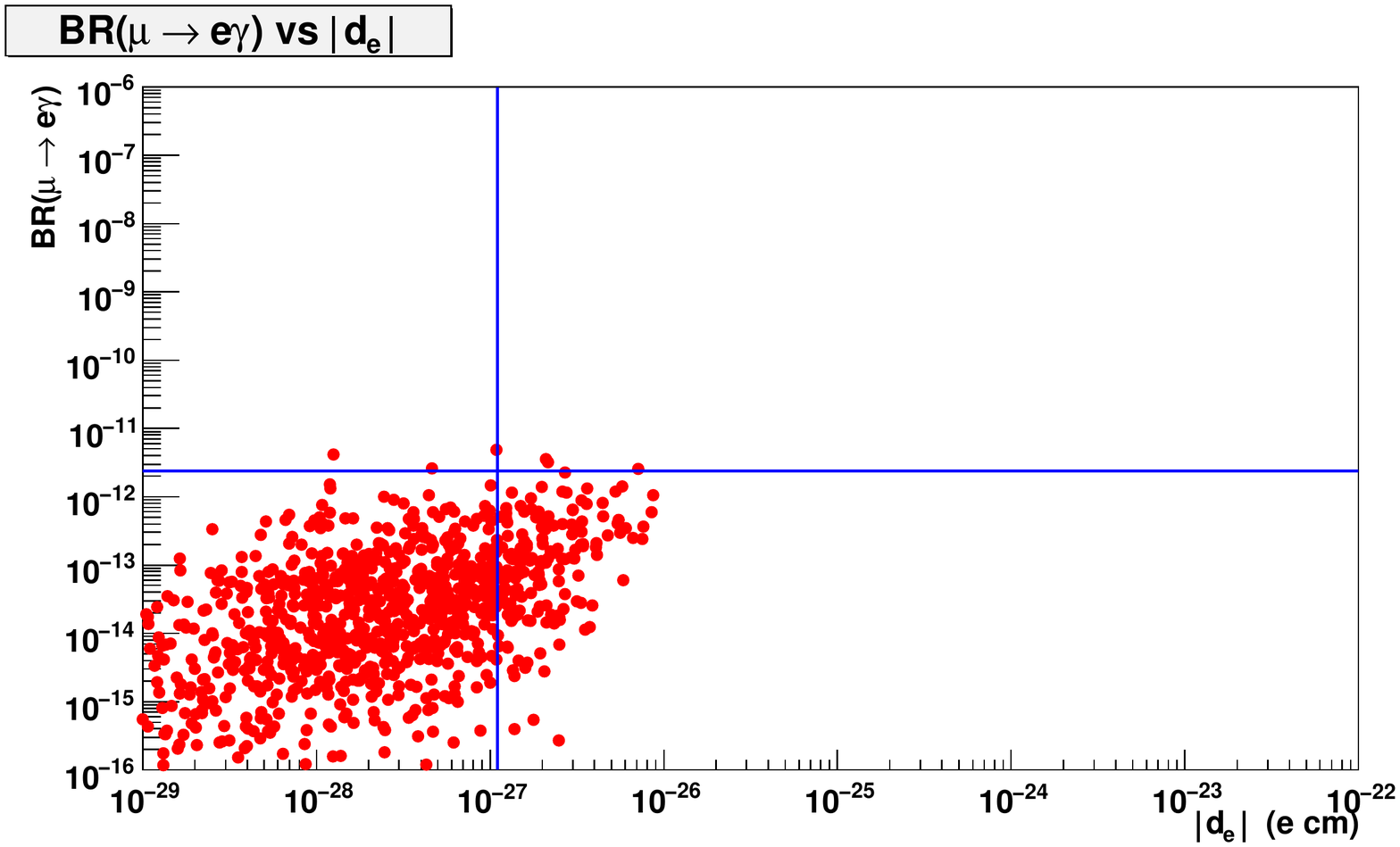}\\
\end{center}
\caption{  $\mbox{BR} (\mu \rightarrow e \gamma)$ vs $\left| d_e \right|$ for
small rotation angles $\theta$. 
Left: $\theta < 0.1$, right: $\theta <0.03$. Here $m_h=200$ GeV.
The experimental limits are given by the straight lines.  }
\label{res2dim6}
\end{figure}

As the next step, we study  constraints on the rotation angles
parametrizing $U_{L,R}$, while
allowing for arbitrary values of the four phases therein as well as
a wide range of proportions between $Y_{ij}^{(0)} $ and 
$Y_{ij}^{(1)} v^2/M^2$ in Eq.~\ref{dim6-Y}.
Specifically, we choose  $Y_{ij}^{(0)}$ in the range $\pm 0.9 \times  Y_{ij}$.
The results are presented in
Fig.~\ref{res2dim6}, where we vary all the angles (in $U_L$ and $U_R$) in the same
range. If one allows for  angles as large as 0.1, both 
  BR$(\mu \rightarrow e \gamma)$ and $d_e$  are overproduced, while the angles
of order $0.03$ are  typically  consistent with the constraints for $m_h=200$ GeV. 
More precisely,  BR$(\mu \rightarrow e \gamma)$ and $d_e$ are most sensitive
to the 1-3 and 2-3 mixing angles  $\theta_{13}, \theta_{23}$  in terms 
of the standard parametrization of unitary matrices \cite{Chau:1984fp}.
We then  find
\begin{equation}
\theta_{13}, \theta_{23} < 3 \times 10^{-2} \times {m_h \over 200~{\rm GeV}} \;,
\end{equation}
while $\theta_{12}$ is allowed to be as large as ${\cal O}(1)$. 
We note that $d_e$ is somewhat more restrictive than BR$(\mu \rightarrow e \gamma)$
and requires $\theta_{13}, \theta_{23} <  10^{-2}$ for $m_h=200$ GeV.

\subsubsection{2--loop constraints}

\begin{figure}
\begin{center}
\hspace{-1.3cm}
\includegraphics[width=8.5cm]{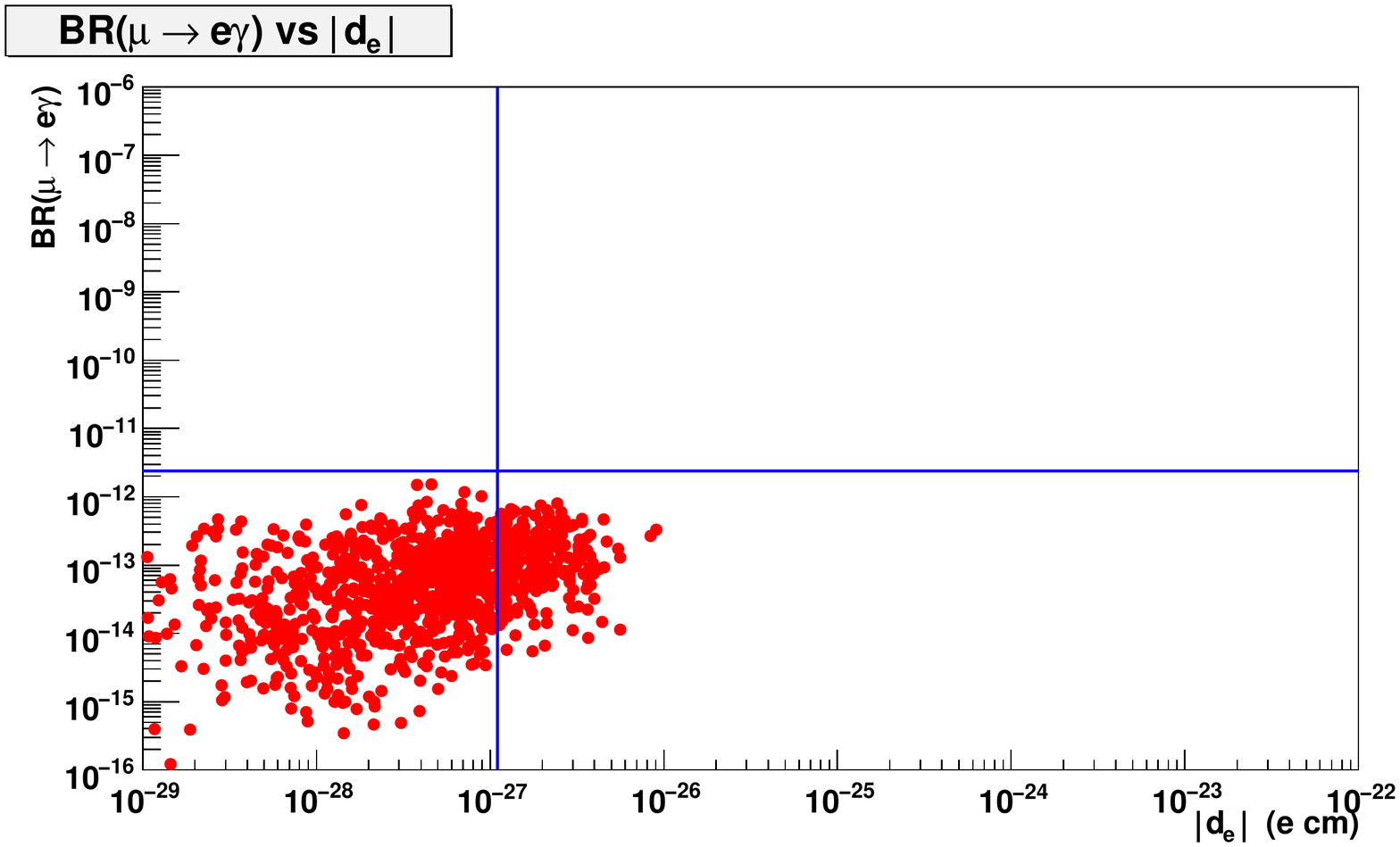}\hspace{0.01cm}
\includegraphics[width=8.5cm]{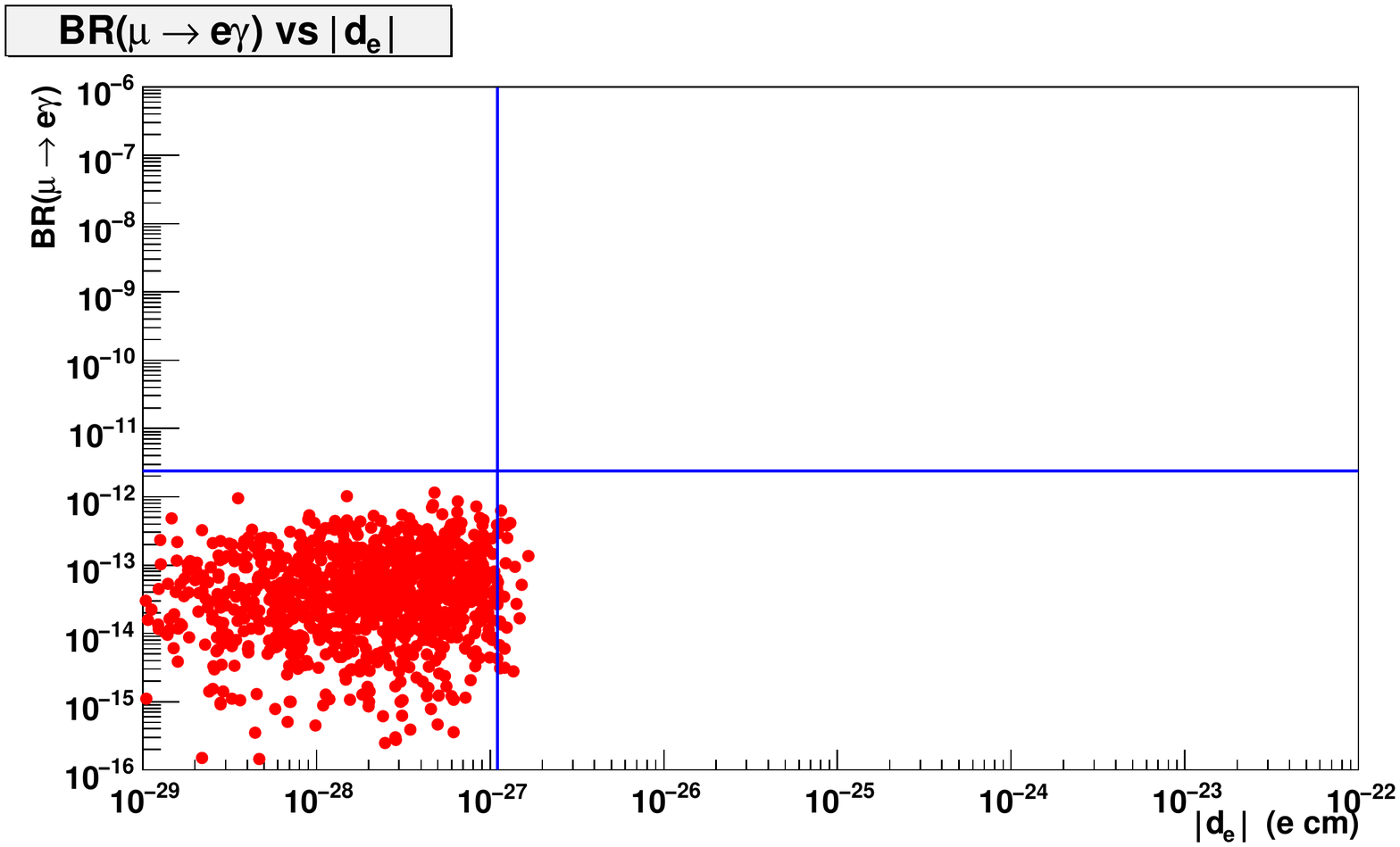}\\
\end{center}
\caption{  $\mbox{BR} (\mu \rightarrow e \gamma)$ vs $\left| d_e \right|$ 
at 2 loops.
Left: the dim--6 contribution is limited to 0.1\%, while the rotation angles are arbitrary. 
Right: the rotation angles are bounded by $\theta_{12}< 10^{-3}, \theta_{13, 23} < 10^{-2}$
and the CP phases $\phi < 0.1$, while the balance between dim--4 and dim--6 contributions
is arbitrary. Here $m_h=200$ GeV.  The experimental limits are given by the straight lines.   }
\label{2loopSCAN}
\end{figure}

The 2--loop contributions are very sensitive to the couplings in the (12)--block. 
In particular, if one assumed a ``natural'' value of $y_{12}$ to be of order 
$\sqrt{m_e m_\mu}/v$,   $\mbox{BR} (\mu \rightarrow e \gamma)$ would be 
overproduced by orders of magnitude, as seen from Table 2. 
The resulting constraints are very strong,
as our numerical study demonstrates in Fig.~\ref{2loopSCAN}.
An analog of Eq.~\ref{const} in the (12)--sector  reads
\begin{equation}
\biggl\vert  Y_{ij}^{(1)} { v^2 \over M^2} \biggr\vert
< 10^{-3}  \vert Y_{ij} \vert  
\label{const2}
\end{equation}
for $m_h=200$ GeV. The Higgs mass dependence is weaker than linear and 
the bound gets tighter by a factor 1.5 for $m_h=100$ GeV.
Interpreting this constraint as is done in Eq.~\ref{interp}, we get
\begin{eqnarray}
   Y_{ij}^{(1)} \sim  Y_{ij} & \Rightarrow & M > 5~{\rm TeV}  ~, \nonumber\\
  Y_{ij}^{(1)} \sim 1 & \Rightarrow & M > 2000~{\rm TeV}  ~.
\label{interp2}
\end{eqnarray}

Allowing for arbitrary proportions between dim--4 and dim--6 operators, 
we find the following constraints on the rotation angles
\begin{equation}
\theta_{12}< 10^{-3} ~~,~~ \theta_{13,23}< 10^{-2} \;.
\end{equation}
Note that $\mbox{BR} (\mu \rightarrow e \gamma)$ is also sensitive to 
$\theta_{13,23}$, although not as much as to $\theta_{12}$.
Unlike in the 1--loop case, a diagonal lepton Yukawa matrix does not solve
all the problems: $d_e$ is sensitive to the phase of the flavor--diagonal coupling 
$y_{11}$ which does not vanish in the limit of zero rotation angles. We thus have
a separate constraint on the CP phase of $y_{11}$:
\begin{equation} 
\phi < 10^{-1} \;.
\end{equation}

\subsection{Higher dimensional operators and the mass hierarchy}

Let us now explore the possibility that the lepton mass hierarchy is entirely
due to higher dimensional operators \cite{Babu:1999me,Giudice:2008uua}. 
The Yukawa couplings are    expanded as
\begin{equation}
Y_{ij} (H)
= \sum_{n_{ij}=0}^\infty \kappa_{ij}^{(n_{ij})}~\left(     
 {H^\dagger H \over M^2 }
\right)^{n_{ij}} \;,
\end{equation}
with order one $ \kappa_{ij}^{(n_{ij})} $ and $M$ being a new physics scale.
In most interesting cases which address the flavor problem, 
the coefficients $ \kappa_{ij}^{(n_{ij})} $ 
vanish up to a certain order $n_{ij}$. 
This can happen
due to some symmetry (e.g. Froggatt--Nielsen type \cite{Froggatt:1978nt}) 
 of the UV completion of our effective theory, 
which may not be apparent at low energies.
In this case, the mass hierarchy is generated by 
\begin{equation}
\epsilon = {v^2 \over M^2} \ll 1 
\end{equation}
and the Yukawa textures take the form 
\begin{equation}
  Y_{ij} = c_{ij} ~\epsilon^{n_{ij}} \;,
\end{equation}
with order one $c_{ij} $. Analogous textures in the quark sector were considered 
in \cite{Giudice:2008uua}. It is interesting that the invariant measure of CP violation
increases by many orders of magnitude compared to that in the Standard Model, which
can be relevant to baryogenesis  \cite{Lebedev:2010zg}.

Various lepton textures of the above type can be generated as follows.
The Yukawa matrix is represented in terms of the eigenvalues and 
the  rotation matrices $U_{L,R}$   as  in Eq.~\ref{eq:biunitary}.
Scanning over $U_{L,R}$ then produces viable lepton textures.
Choosing for definitness $\epsilon =1/60$ as 
motivated by the top--bottom quark mass hierarchy \cite{Giudice:2008uua},
one determines the exponents $n_{ij}$ via
\begin{equation}
  n_{ij} = \mathrm{round}\!\left( \log_\epsilon |Y_{ij}| \right) \;,
\end{equation} 
which also fixes the ${\cal O}(1)$ coefficients $c_{ij}$.
Moving to the mass eigenstate basis, we obtain the lepton
couplings of the physical Higgs (\ref{Lhiggs}),
\begin{equation}
  y_{ij} = (U_L)_{ki}^* \; (2n_{kl} + 1) Y_{kl} \; (U_R)_{lj} \;.
\label{yij}
\end{equation}
Note that, in this basis,  $y_{ij}$ 
are defined up to a phase (see \cite{Lebedev:2010zg}
for details). Indeed, one can
multiply the left-- and right--handed leptons by diagonal
phase matrices such that the mass terms remain the same. 
However, observables are invariant under this reparametrization
since they involve combinations like $y_{3i}y_{i3}$ or
absolute values of the couplings.

For concreteness, let us restrict ourselves to Yukawa textures of the following type
\begin{eqnarray}
Y \sim  \left(
\begin{matrix}
\epsilon^3 & \epsilon^l  & \epsilon^m \\
\epsilon^l & \epsilon^2  & \epsilon^n \\
\epsilon^m & \epsilon^n  & \epsilon^1 
\end{matrix}
\right) \;. \label{texture}
\end{eqnarray}
Here we have chosen the diagonal entries such that they reproduce (roughly)
the $m_e : m_\mu : m_\tau $ hierarchy. The integers $l,m,n$ are subject to
the LFV and CP constraints. For large $l,m,n$ the Yukawa matrix is approximately 
diagonal and the constraints are satisfied. Numerically, we find 
 the following approximate bound:
\begin{equation}
l\geq 4 ~~,~~ m\geq 2 ~~,~~ n \geq 2 \;.
\end{equation}
For most choices of order one coefficients in the texture, the constraints
are satisfied in this case. Note that the bound on  $l$ is the strongest one
due to the Barr--Zee contributions to $\mu \rightarrow e \gamma$. 
This entails, in particular, that the lepton  analog of the down quark texture studied
in Ref.~\cite{Giudice:2008uua}  is strongly disfavored.

\begin{figure}
  \centering
  \subfigure[]{\includegraphics[height=16em]{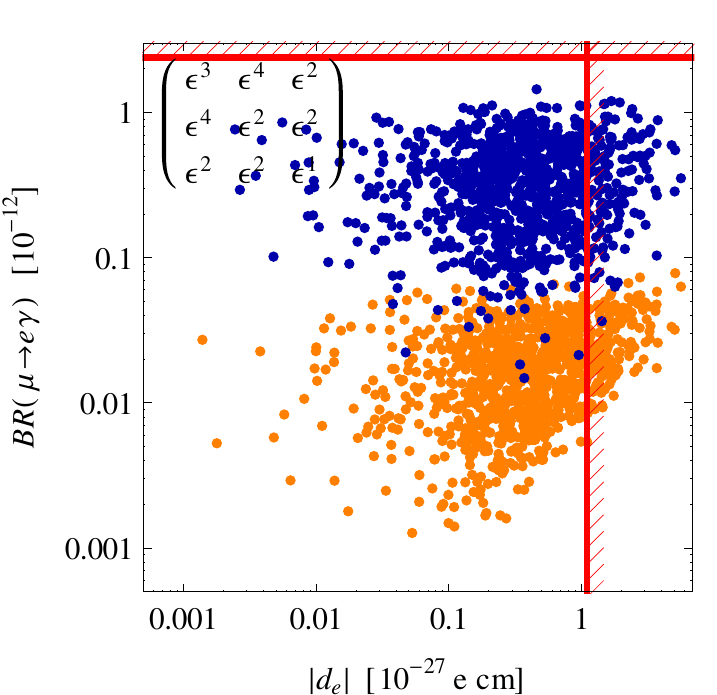}}
  \subfigure[]{\includegraphics[height=16em]{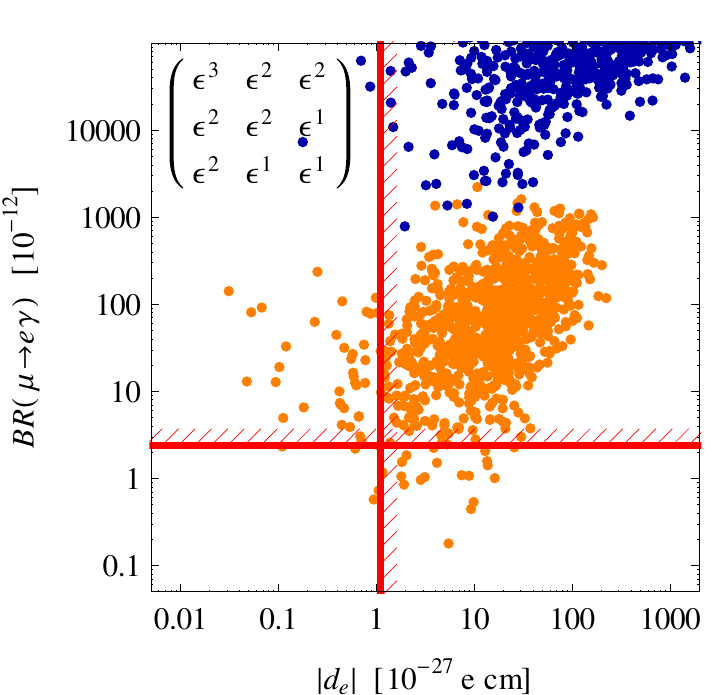}}
  \caption{BR$(\mu\rightarrow e\gamma)$ vs $|d_e|$
    for representative Yukawa textures,
    with $m_h = 200\ \mathrm{GeV}$. Light (orange) and dark (blue) dots
correspond to one loop and one + two loop  contributions, respectively.
 The experimental limits are given by the straight lines.    }
  \label{fig:meg vs de}
\end{figure}

Our numerical results for representative textures are shown in Fig.~\ref{fig:meg vs de}.
We take $l=4$, $m=n=2$ for panel (a) and $l=m=2$, $n=1$ for panel (b). 
In this figure,  the light (orange) and dark (blue) dots
correspond to one loop and one loop plus  two loop  contributions, respectively.
For case (a) most points satisfy the constraints, whereas in case (b) the bounds are grossly 
violated. We find that BR$(\mu\rightarrow e\gamma)$ is dominated by the Barr--Zee contributions,
whereas $d_e$ receives comparable contributions at 1 and 2 loops.
As expected, significant departures from the diagonal Yukawa form,
especially in the (12)--block, are not allowed.

Finally, let us note that although the constraints are strong, the remarkable 
sensitivity of  $\mu\rightarrow e\gamma $ and $d_e$ also implies good 
prospects for detection of Higgs--induced lepton flavor and CP violation
in  current and future experiments.

\section{Conclusion}

In this work, we have derived model--independent  bounds on the 
lepton  flavor violating  couplings of the SM Higgs boson.
Such flavor violation appears when higher dimensional operators 
contribute to the lepton masses. Scanning over various textures shows
that these contributions should be limited to about 0.1\% in the (12)--sector
and no more than 10\% in the (13) and   (23)--sectors, otherwise
BR($\mu \rightarrow e \gamma$) and $d_e$ are overproduced. 
Alternatively, if one allows for large contributions
of the higher order operators, the Yukawa matrix must be diagonalizable
by a small--angle rotation with $\theta_{12} \sim {\cal O} (10^{-3} ) $ and 
$\theta_{13}, \theta_{23} \sim {\cal O} (10^{-2} )$. In addition, the electron 
EDM sets a constraint on the relevant 
CP phase $\phi < {\cal O} (10^{-1} )$. 

Further, we have  studied the possibility that the lepton mass hierarchy
is created entirely by non--renormalizable operators.  Also in this case the
LFV effects are significant. The preferred textures have small intergenerational
mixing, especially in the (12)--sector, as required by the Barr--Zee 
contributions to $\mu \rightarrow e \gamma$. As a result, the lepton
analog of the down quark texture studied in \cite{Giudice:2008uua} is
strongly disfavored.

On the other hand, the remarkable sensitivity of LFV and CP violating
observables to Higgs--induced effects implies good experimental
prospects for detection of $\mu \rightarrow e \gamma$ and $d_e$.

\section*{Acknowledgments}

We are grateful to S. Davidson for useful comments.
The work of A.G. is supported in part by the
Landes-Exzellenzinitiative Hamburg. J.P. is supported in part by
German Research Foundation DFG through
Grant No.\ STO876/2--1.

\end{document}